\documentclass[12pt,doublespace]{article}
\begin{document}
\begin{center}
CONTROLLING CHAOS IN SOME LASER SYSTEMS VIA VARIABLE COUPLING AND FEEDBACK TIME DELAY\\
E. M. Shahverdiev $^{1}$ \\
Institute of Physics, H. Javid Avenue, 33, Baku, AZ1143, Azerbaijan\\
$^{1}$e-mail: shahverdiev@physics.ab.az\\
ABSTRACT
\end{center}
We study numerically a system of two lasers cross-coupled optoelectronically with a time delay where the output intensity of each laser modulates the pump current of the other laser. We demonstrate control of chaos via variable coupling time delay by converting the laser intensity chaos to the steady state. We also show that wavelength chaos in an electrically tunable distributed Bragg reflector laser diode with a feedback loop can be controlled via variable feedback time delay. 
~\\
\begin{center}
I INTRODUCTION
\end{center}
\indent Time-delayed systems are ubiquitous in nature, technology and society because of finite signal transmission times, switching speeds and memory effects [1-5]. Because of their ability to generate high-dimensional chaos, these systems are good candidates for chaos based secure communication. In this context laser systems with a time delay are of considerable practical significance [3-7].\\
\indent Usually time delay(s) are considered as a chaos creator in otherwise non-chaotic systems. Consider for example one of the paradigmatic and most studied models in chaos theory-one dimensional Ikeda model: $dx/dt=-\alpha x +\beta \sin(x-\tau)$. Physically $x$ is the phase lag of the electric field across the resonator; $\alpha $ is the relaxation coefficient; $\beta$ is the laser intensity injected into the system; $\tau $ is the round-trip time of the light in the resonator. The Ikeda model plays an important role in electronics and physiological studies. This model was introduced to describe the dynamics of an optical bistable resonator and is well known for delay induced chaotic behavior, see [8-12] and references therein. It is clear that for $\tau=0$ there is no chaotic dynamics for this model, as the Poincar{\'e}-Bendixson theorem [13] requires that autonomous first-order ordinary differential equations with continuous functions be at least three dimensional to have chaotic solutions [14]. For $\tau>0$ time delay may generate instability and make the system chaotic even for the simplest systems, as the delay time renders the Ikeda system's phase space infinite dimensional. It is also noted that the Ikeda model or its modifications can be used for the description of the dynamics of some laser systems such as class B lasers [15-16]. Typical representatives of class B systems are the solid-state, semiconductor, and low pressure $CO_{2}$ lasers [16].\\
\indent Other typical examples of delay-induced chaotic behavior include laser systems with feedback(s)[6-7]. One famous example in the chaos theory and laser physics is the Lang-Kobayashi model with feedback [17-20]. Without feedback the laser relaxes to a constant intensity. However for already very weak feedback, the laser is unstable and its intensity becomes chaotic, see [18] and references therein.\\
\indent Due to the finite velocity of signal transmission between the interacting systems, the couplings also might be delayed. Such a delay can cause chaotic dynamics in the coupled systems [21]. Again consider two Ikeda model coupled bidirectionally: $dx/dt=-\alpha x +\beta \sin x + \beta_{1}\sin(y-\tau_{1})$, $dy/dt=-\alpha y +\beta \sin y + \beta_{2}\sin(x-\tau_{1})$. Here $\beta_{1,2}$ is the coupling strength; $\tau_{1}$ is the coupling time delay. Without the time delay $\tau_{1}=0$ we have two first order ordinary differential equations. For above mentioned reason such a system can not display chaotic dynamics. However, with the time delay the system of two Ikeda models becomes infinite dimensional and there is a possibility that its dynamics could become chaotic.\\
\indent Further, for the optoelectronically coupled laser systems with parameter values used in numerical simulations in this paper without the coupling delay the laser intensity goes to the steady state level. As demonstrated below the coupling time delay can generate a chaotic dynamics for the laser intensity.\\   
\indent Following [22] in [23] we have studied a system of two lasers cross-coupled optoelectronically with a time delay where the output intensity of each laser modulates the pump current of the other laser. As emphasized in [22], these systems are analogous to the equations describing the spread of certain diseases such as dengue epidemics in human populations coupled by migration, transportation, etc. In-phase oscillations (complete synchronization) between such epidemic models, which favor the spread of diseases, are considered in [24]. In [23] we have considered the possibility of both the in-phase (complete) and anti-phase (inverse) chaos synchronization between the cross-coupled laser models. We have established that by changing the coupling strength between the systems the transition from the in-phase(complete) synchronization to the anti-phase (inverse) synchronization might be realized. The results can be important for the disruption of the spread of the certain infectious diseases in human populations. Indeed, as we have established in [23] for higher values of the coupling strength between the interacting systems (say two areas of the epidemic infection region) in-phase synchronization occurs. Such in-phase synchronization can increase the danger of further epidemic spreading. Reducing the coupling strength we have found that there is an anti-phase synchronization between the systems. Such an anti-phase synchronization could prevent the spread of disease. The coupling strength can be reduced via changes in the connectivity between the systems by governing migration, transportation, etc. Quite interestingly there is no need to make the coupling strength between the interacting systems negligibly small: in [23] we have found that in-phase synchronization occurs for the coupling strength K=6, anti-phase synchronization- for K=2.06. It should be noted that the role of the in-phase synchronization and anti- phase synchronization in relation to the global extinction of species was underlined earlier in [25].\\
\indent Control of chaos refers to a process wherein a perturbation is applied to a chaotic system, in order to realize a desirable (chaotic, periodic, or stationary) behavior [26]. The idea of chaos control was enunciated by E. Ott, C. Grebogi and J. Yorke (OGY) in 1990 in [27], where a method for stabilizing an unstable periodic orbit was suggested. The main idea consisted in waiting for a natural passage of the chaotic orbit close to the desired periodic behavior, and then applying a small judiciously chosen perturbation, in order to stabilize such periodic dynamics. In [28] Pyragas suggested a feedback mechanism in which a state variable is directly perturbed such as to control a periodic orbit. In such a case, a feedback term is proportional to the difference between the actual value of the state variable, and the value delayed of a time lag. The idea is that, when the time lag coincides with the period of one unstable periodic orbit of the unperturbed system, the feedback pushes to zero the difference between the present and the delayed dynamics, and the periodic 
orbit is stabilized. Ideas by OGY and Pyragas initiated an avalanche of research in the field of chaos control and synchronization control, see e.g. [29-39] and references therein.\\ 
\indent In the field of chaos control most attention in the literature is given to the study of the role of time-delayed feedback. In particular, in [36] adaptive modification of the delayed feedback control algorithm with a continuously varying time delay was proposed. Research on the delay-coupled systems are shown to lead to many interesting phenomena such as oscillation death, stabilizing periodic orbits, enhancement or suppression of synchronization, chimera state, etc. [40]. There is also a lot of research where the coupling strength ({\it not the coupling time delay}) is modified to control 
chaos in delay-coupled systems and synchronization control in delay-coupled networks, see e.g.[37-38] and references therein. However, the research on the role of the variable coupling and feedback time delay in controlling chaos in real world systems is relatively scarce. Some references can be found in [40-41].\\
\indent In this paper we first demonstrate chaos control in cross-coupled laser systems via variable coupling time delay. We show that the steady state (stationary) behavior can be realized from the chaotic state. Then we also study the dynamics of wavelength chaos with the variable feedback time delay. In both the cases we consider different types of the coupling and feedback time delay: the sinusoidal time delay, the chaotic time delay and the combined sinusoidal and chaotic time delay. To the best of our knowledge such a detailed approach to chaos control with coupling and feedback time delay is proposed for the first time. We establish that for the case of variable coupling time delay for chosen parameter values all the three modulations of the time delay result in the steady state from the chaotic dynamics and in terms of the transition time the combined sinusoidal and chaotic time delay scenario performs better.\\
\indent In the paper we also investigate wavelength chaos dynamics in an electrically tunable distributed Bragg reflector laser diode with a feedback loop via variable feedback time delay. In the case of variable feedback time delay we achieve the steady state level only with the combined sinusoidal and chaotic time delay scenario. Other two scenarios fail to convert the chaotic wavelength to the steady state. These results can be of certain importance for obtaining stable laser sources in real world situations.\\
\indent The paper is organized as follows. In Section II we introduce the model of delay-coupled lasers. Numerical simulation of the model and discussion of the results are presented in Section III. Section IV draws conclusions.
\begin{center}
II ORIGINAL LASER MODEL AND DELAY-COUPLED LASERS
\end{center}
\indent Following [22] we consider a standard semiconductor laser model with the following intensity and carrier number dynamics
\begin{equation}
\frac{dI}{dt}=gIN -\tau_{p}^{-1}I, 
\end{equation} 
\begin{equation}
\frac{dN}{dt}=P - gIN -\tau_{c}^{-1}N, 
\end{equation} 
where $I$ is the photon number and describes fast dynamical processes; $N$-slow dynamical variable is the carrier number;  Source term $P$ is the pump rate; $g$ is the gain coefficient responsible for the nonlinear coupling between $I$ and $N$; $\tau_{p}$ is the photon lifetime; $\tau_{c}$ is the carrier lifetime.\\
\indent By introducing new variables $J,D$ and $t_{1}$
\begin{equation}
I = I_{0}J, N=N_{0}D, t=t_{0}t_{1},    
\end{equation} 
with
\begin{equation}
t_{0}=\tau_{p} , N_{0}=\tau_{p}^{-1}g^{-1}, I_{0}=\tau_{c}^{-1} g^{-1},  
\end{equation} 
and
\begin{equation}
\epsilon=\tau_{p}\tau_{c}^{-1} ,A=PN_{0}^{-1}\tau_{c},  
\end{equation} 
the original laser model (1) and (2) can be rewritten as:
\begin{equation}
\frac{dJ}{dt_{1}}=(D-1)J, 
\end{equation} 
\begin{equation}
\frac{dD}{dt_{1}}=\epsilon^{2}(A-(J+1)D). 
\end{equation} 
With a sequence of transformations (for some details see [43]), the system of Eqs. (6) and (7) can be rescaled to  Eqs. (8) and (9)
\begin{equation}
\frac{dy}{dt}=x(1+y), 
\end{equation} 
\begin{equation}
\frac{dx}{dt}=-y -\epsilon x(a + by) 
\end{equation} 
with 
\begin{equation}
y=(J-J_{0})J_{0}^{-1},x=(D-D_{0})\epsilon^{-1}J_{0}^{-0.5},a=(1+J_{0})J_{0}^{-0.5},b=J_{0}^{0.5},t=\epsilon J_{0}^{0.5}t_{1}
\end{equation} 
We note that $D_{0}=1$ and $J_{0}=A-1$ are the steady states for Eqs. (6) and (7).It is also noticed that $y$ is the intensity fluctuation normalized about the steady state level.\\
\indent Now consider two cross-coupled laser systems labeled with the indices 1 and 2,
\begin{equation}
\frac{dy_{1}}{dt}=x_{1}(1+y_{1}), 
\end{equation}
\begin{equation}
\frac{dx_{1}}{dt}=-y_{1} -\epsilon x_{1}(a + by_{1})-\epsilon K_{2}y_{2}(t-\tau),
\end{equation} 
and 
\begin{equation}
\frac{dy_{2}}{dt}=x_{2}(1+y_{2}), 
\end{equation} 
\begin{equation}
\frac{dx_{2}}{dt}=-y_{2} -\epsilon x_{2}(a + by_{2})-\epsilon K_{1}y_{1}(t-\tau).
\end{equation} 
Where $\tau$ is the coupling time delay and $K_{1}$  and  $K_{2}$ are the coupling strengths. 
\begin{center}
III NUMERICAL SIMULATIONS AND DISCUSSION
\end{center}
\indent Now we present numerical simulation results for the optoelectronically cross-coupled laser systems, Eqs. (11-14). In the numerical simulations we choose the following values for the main parameters appropriate to semiconductor lasers [22]:$\epsilon=(0.001)^{0.5}, a=2, b=2.33$. Coupling  constants are $K_{1}=4$  and  $K_{2}=4.$ First we  consider the case of a constant coupling time delay $\tau =30.$ Fig. 1 shows chaotic dynamics for the intensity of laser $y_{1}(t)$, described by Eqs.(11-12).\\
\indent Next let us consider the variable time delay scenarios. We have experimented with different types of variable time delay: (a) sinusoidal modulation of the time delay; (b) chaotic modulation of the time delay; and (c) combined sinusoidal and chaotic modulation of the time delay.\\
For sinusoidal modulation we take $\tau(t)=\tau +\tau_{a}\sin\omega t$, where $\tau$ is the zero-frequency component 
(a constant time delay), $\tau_{a}$ is the amplitude,$\omega$ is the frequency of the modulation. In Fig.2 we present the intensity dynamics of the laser $y_{1}(t)$ for $\tau (t)=30 + 20\sin t$. It demonstrates that after some transient the intensity fluctuations tend to zero, in other words the intensity of the laser $y_{1}(t)$ approaches steady state level (see Eq. (10)). Fig.3 demonstrates the case of 
the laser intensity approaching the steady state level for $\tau (t)=30 + 20\lambda(t).$ Where $\lambda(t)$ describes the wavelength chaos dynamics according to Eq. (15) 
\begin{equation}
\frac{d\lambda(t)}{dt}=-\alpha \lambda(t) + m \sin^{2}(\lambda(t-\tau) - \Phi_{0})
\end{equation} 
where $\alpha$ is the relaxation coefficient, $m$ is the feedback strength in the electooptic laser model (for more details, see e. g. [44] and references therein) and $\Phi_{0}$ is the feedback phase. The laser system considered in Eq. (15) is an electrically tunable Distributed Bragg Reflector (DBR) laser diode with a feedback loop. This system was proposed in [44] as a chaotic wavelength signal generator for chaos based secure communication. For $\alpha=4, m=20, \tau=5$ the system (15) exhibits chaotic dynamics; we also choose $\Phi_{0}=\pi/4$ (Fig.4).\\
Finally, we consider the combined effect of sinusoidal and chaotic modulation of the time delay $\tau (t)=30 + 20\lambda(t)\sin t$ on the laser $y_{1}(t)$ dynamics. Fig.5 depicts $y_{1}(t)$ dynamics for this case. It is seen that in the case of combined time delay modulation, in comparison with the cases (a) and (b), transition from the chaotic state to the stationary one can be shortened.\\
\indent So far we have considered the effect of the {\it variable coupling time delay} on the dynamics of the laser intensity. More precisely, we have shown that variable coupling time delay can be helpful in achieving the steady state for the laser intensity. It appears that {\it variable feedback time delay} can also play an important role in converting the chaotic dynamics of the wavelength to the steady state in the laser systems with a feedback loop. For this purpose we have studied the dynamics of the wavelength chaos for DBR laser diode with a feedback loop, Eq. (15). For numerical simulations we choose the following parameter values: $\alpha=1, m=16, \Phi_{0}=\pi/4 $. It should be noted that although the systems investigated are of the different nature, there are significant differences between the effects of the variable coupling time delay and variable feedback time delay on the systems' dynamics. Fig.6 depicts dynamics of the wavelength for the cases of a constant (solid line) time delay ($\tau=3$) and sinusoidal modulation of the feedback time delay ($\tau (t)=3 + 2\sin 10t $,dotted line). In Fig.7 it is demonstrated that how wavelength dynamics behaves under the effect of a constant ($\tau=3$, solid line) and the chaotic ($\tau (t)=3 + 2\lambda(t)$,dotted line) feedback time delay. Here $\lambda(t)$ is the solution of Eq. (15) for parameter values $\alpha=1, m=16, \tau=3, \Phi_{0}=\pi/4 $. As follows from these two cases, both sinusoidal and chaotic modulation of the feedback time delay fail to convert chaotic wavelength dynamics to the steady state level. As shown by the numerical simulation combined effect of the sinusoidal and chaotic modulation of the feedback time delay succeeded in the controlling chaotic behavior. Fig.8 shows the dynamics of the wavelength corresponding to the case of a constant ($\tau=3$) and modulated in a combined manner feedback time delay: $\tau(t)=3 + 2\lambda(t)\sin(10t)$. It is clear that combined modulation of the feedback time delay outperforms the two other cases in the steering wavelength chaotic dynamics to the steady state.\\
It is also noted that modulation of the time delay can be achieved by a vibrating mirror, e.g. with a piezoelectric transducer-driven mirror [45].
\begin{center}
IV CONCLUSIONS 
\end{center}
\indent In the paper we have demonstrated chaos control in the optoelectronically cross-coupled laser systems via variable coupling time delay. We showed that variable coupling time delay can convert the chaotic intensity dynamics to the steady state. We have also investigated the effect of the variable feedback time delay on the dynamics of the wavelength chaos. In studying chaos control we have considered different types of the coupling and feedback time delay: sinusoidal time delay, chaotic time delay and combined sinusoidal and chaotic time delay. It is established that for the case of variable coupling time delay for chosen parameter values all the three types of modulation of the coupling time delay succeed in converting the chaotic laser intensity to the steady state. It should be noted that in terms of the transition time the combined sinusoidal and chaotic time delay scenario performs better than other scenarios.\\
\indent In the case of variable feedback time delay we have achieved steady state level for the wavelength only with the combined sinusoidal and chaotic time delay scenario. Other scenarios failed to convert chaotic wavelength to the stationary state.\\
The obtained results show that under some circumstances variable coupling and feedback time delay can play the role of chaos eliminator. These results can be of certain importance for obtaining stable laser sources in terms of the intensity and the wavelength. In a wider sense the results of the paper are important from the point of view of chaos control in the interdisciplinary non-linear dynamics.\\
\newpage
\begin{center}
FIGURE CAPTIONS
\end{center}
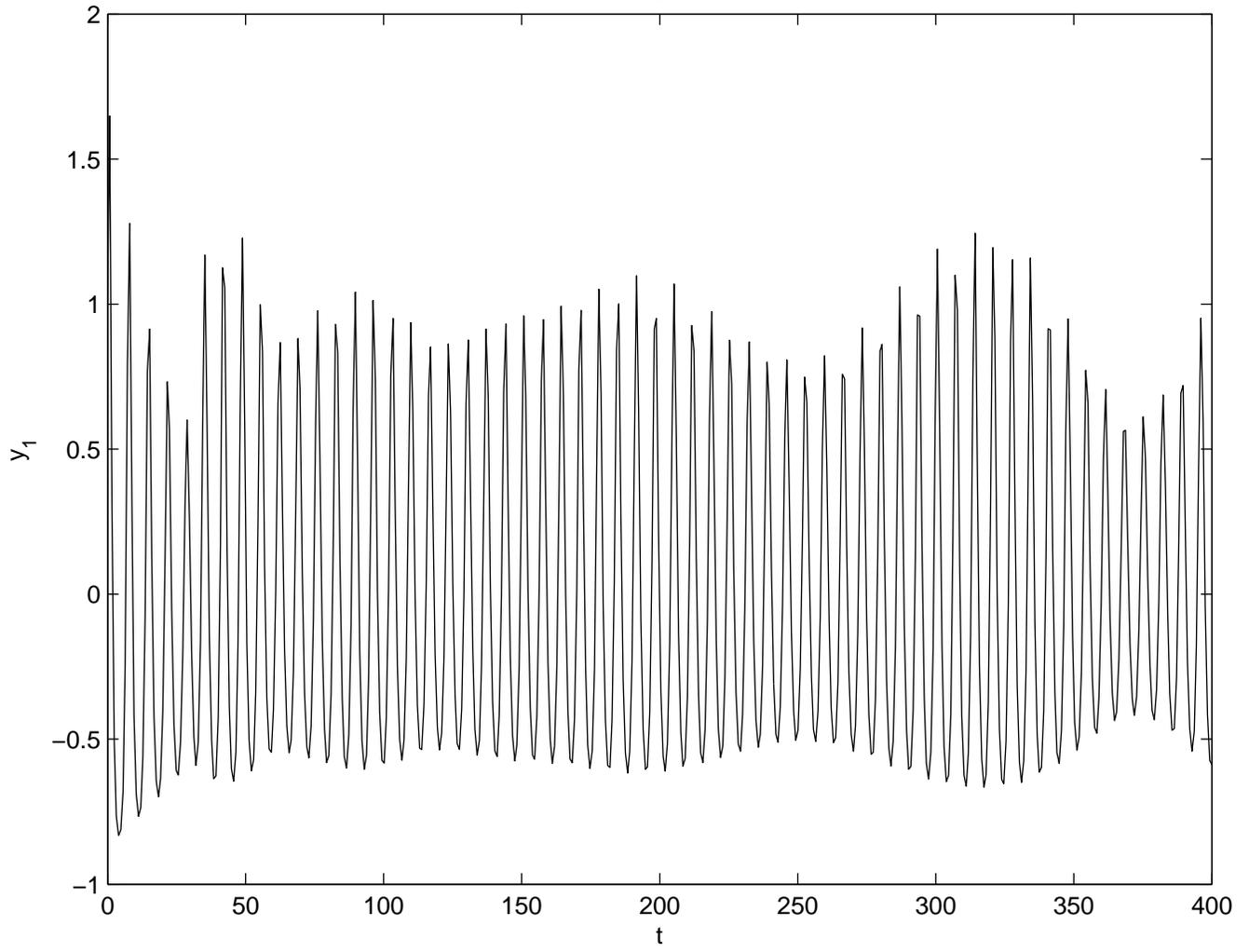
\noindent FIG.1. Numerical simulation of the cross-coupled lasers, Eqs. (11-14): The intensity time series of the laser $y_{1}(t)$ for $\epsilon=(0.001)^{0.5}, a=2, b=2.33, K_{1}=4,$  and  $K_{2}=4$. The coupling time delay between the lasers is a constant: $\tau =30$. Dimensionless units.\\
~\\
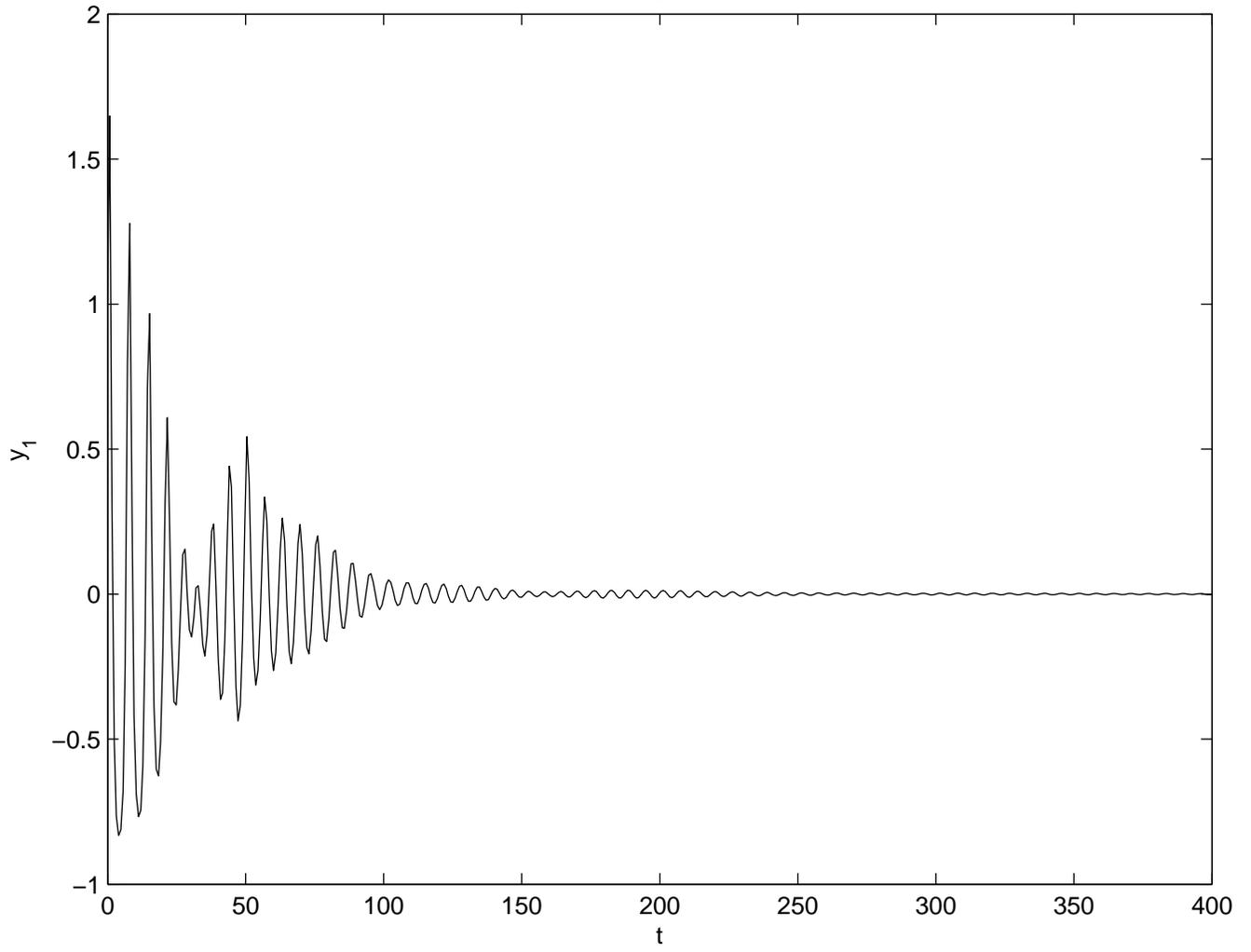
FIG.2. The time series of the laser intensity $y_{1}(t)$ for the case of the sinusoidal coupling time delay between the cross-coupled lasers, Eqs. (11-14): $\tau (t)=30 + 20\sin t$. Parameter values as in Fig.1. Dimensionless units. It is evident that after some transients the laser intensity tends to the steady state level.\\
~\\
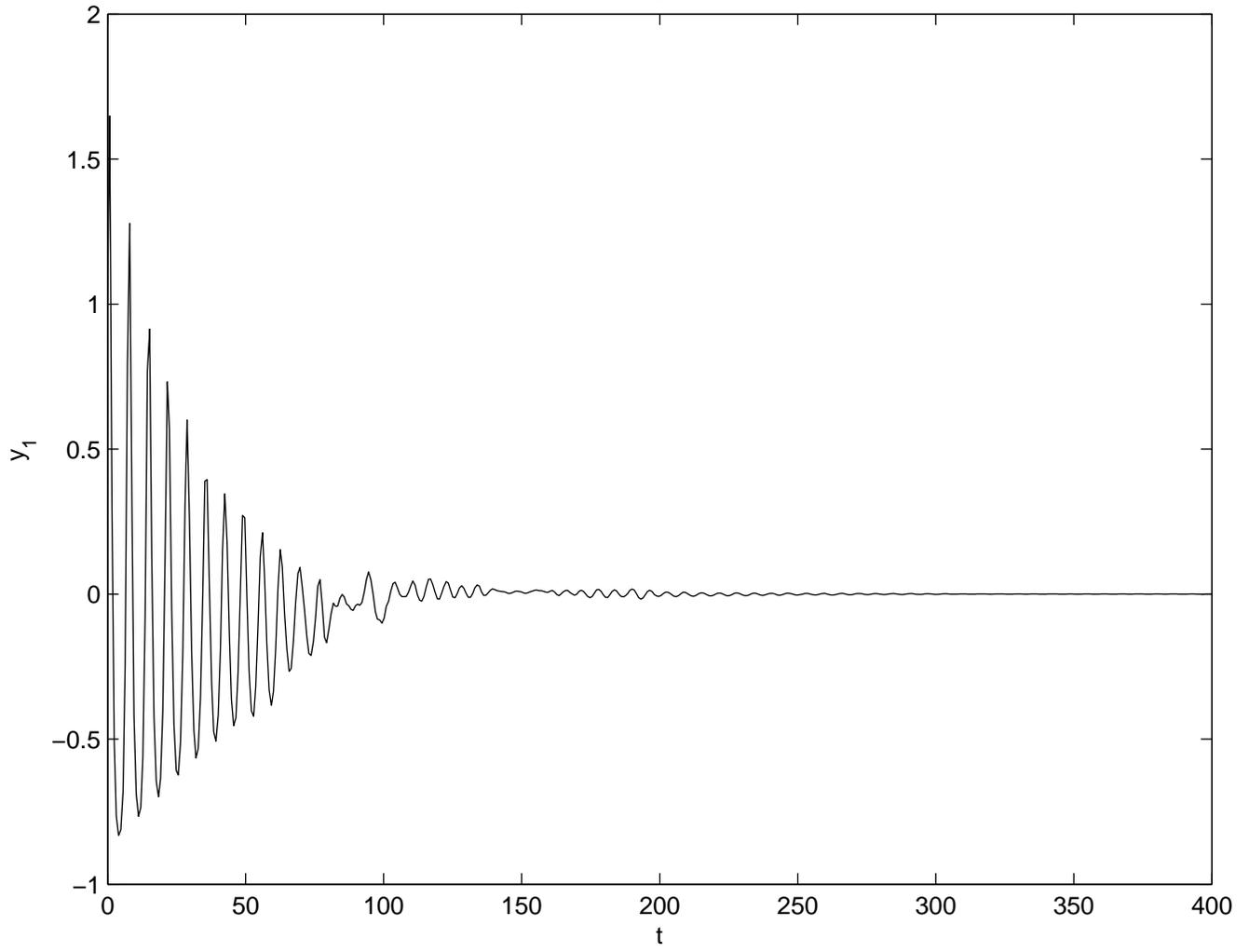
FIG.3. The laser intensity $y_{1}(t)$-time plot for the case of the chaotic coupling time delay between the cross-coupled lasers, Eqs. (11-14): $\tau (t)=30 + 20\lambda(t)$ for $\epsilon=(0.001)^{0.5}, a=2, b=2.33, K_{1}=4,$ and $K_{2}=4$. Dimensionless units. The chaotic laser intensity is steered towards the steady state via the chaotic coupling time delay.\\
~\\
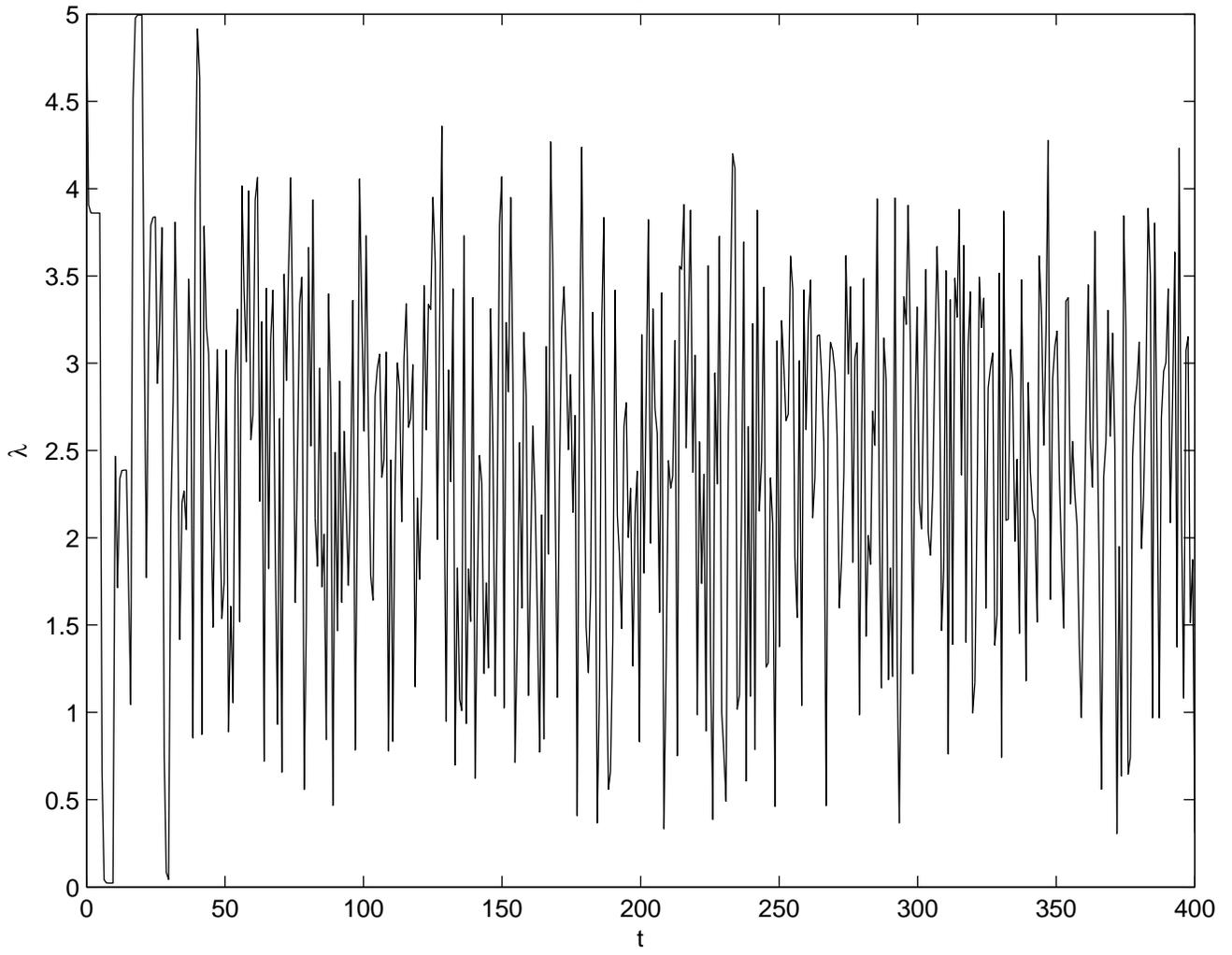
FIG.4. The wavelength $\lambda$ time series of the electrooptical lasers (Eqs. (15)) for $\alpha=4, m=20, \tau=5, \Phi_{0}=\pi/4$. Dimensionless units.\\
~\\
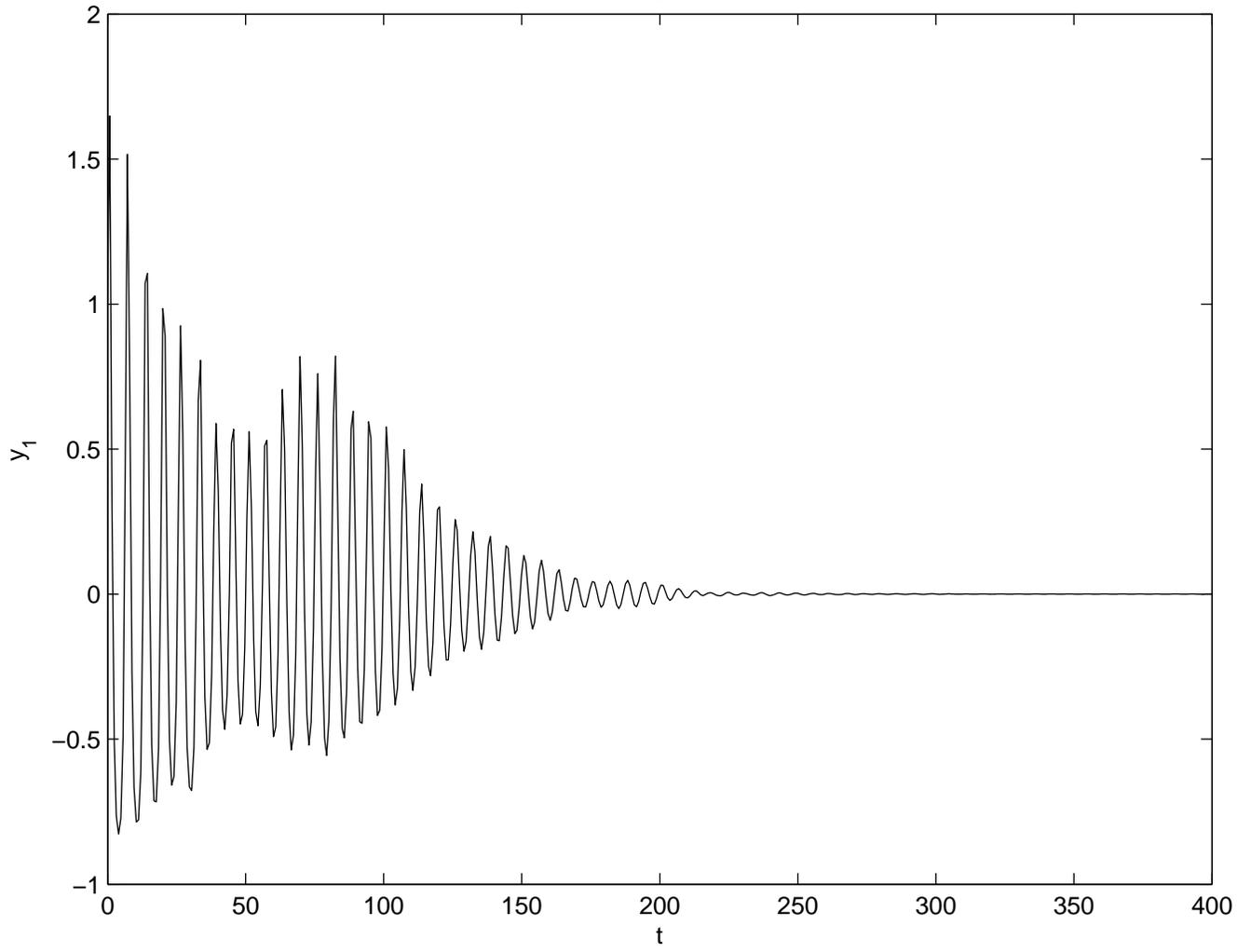
FIG.5. The laser intensity $y_{1}(t)$ versus time $t$ for the case of the combined sinusoidal and chaotic coupling time delay: $\tau (t)=30 + 20\lambda(t)\sin t$ for parameter values as in Figs.1 and 4. Dimensionless units. It is clear that after some transients, due to the effect of the combined sinusoidal and chaotic coupling time delay, the laser intensity approaches the steady state level.\\
~\\
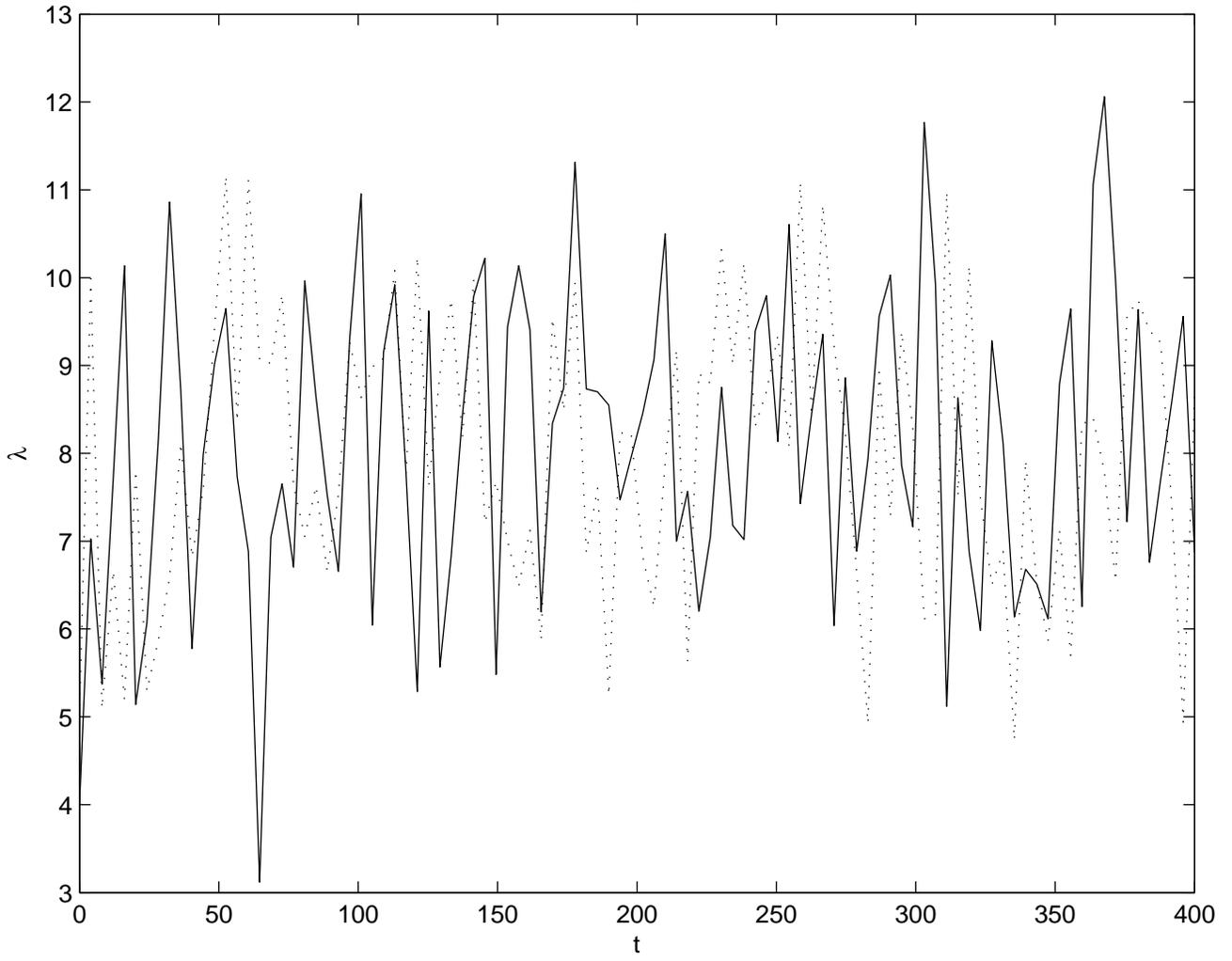
FIG.6. Numerical simulation of the wavelength chaos model, Eq. (15) for the sinusoidal time 
delay feedback: $\tau (t)=3 + 2\sin 10t$. The time series of the wavelength for the variable (dotted line) and a constant (solid line) time delay are shown. Parameter values are: $\alpha=1, m=16, \Phi_{0}=\pi/4 $. Dimensionless units. Sinusoidal time delay feedback fails to convert wavelength chaos to the stationary state.\\
~\\
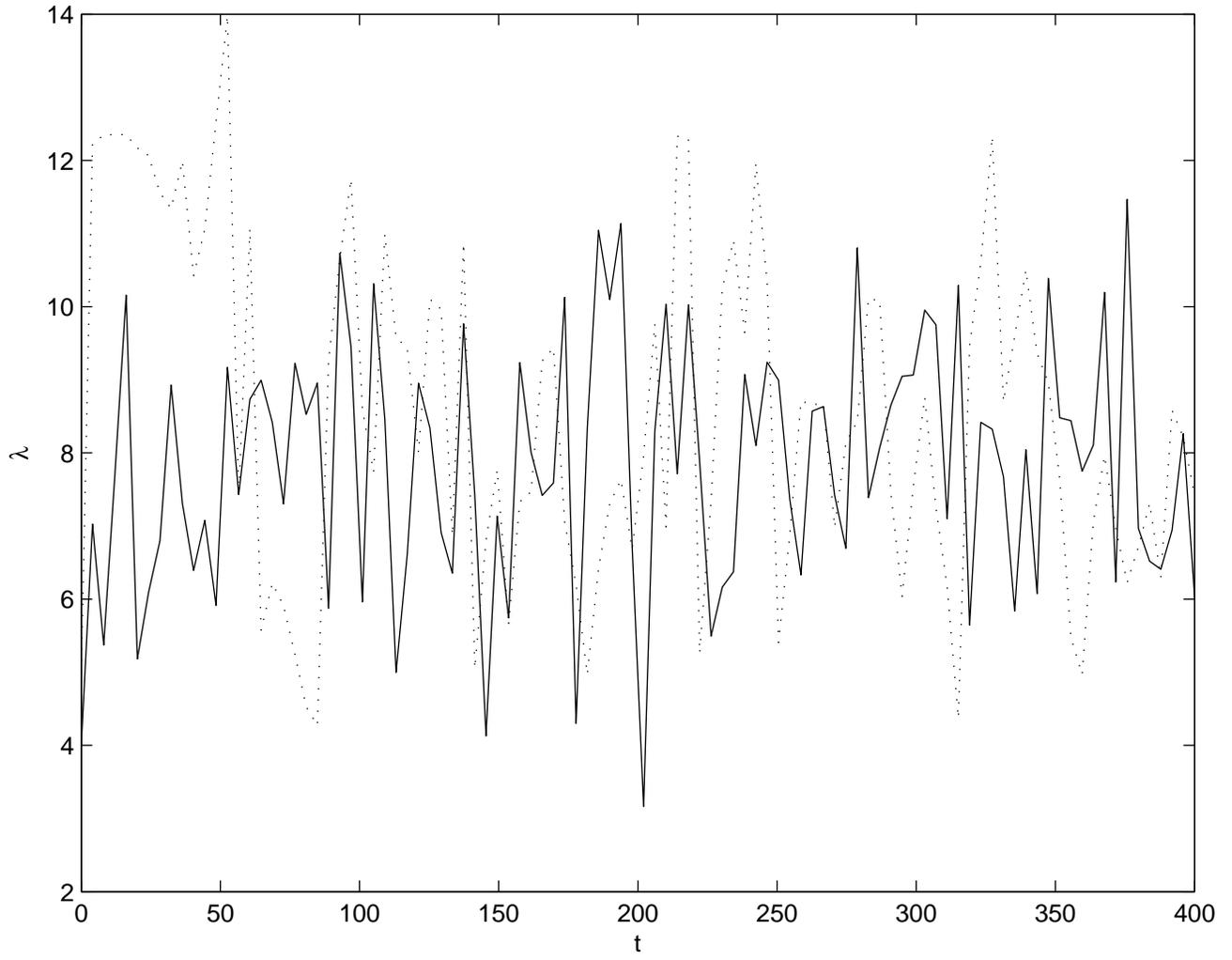
FIG.7. Numerical simulation of the wavelength chaos model, Eq. (15) for the chaotic time 
delay feedback: $\tau (t)=3 + 2\lambda(t)$. The time series of the wavelength for the variable (dotted line) and a constant (solid line) times delay are shown. Parameter values as in Fig.6. Dimensionless units. It is evident that chaotic time delay feedback fails to convert wavelength chaos to the steady state.\\
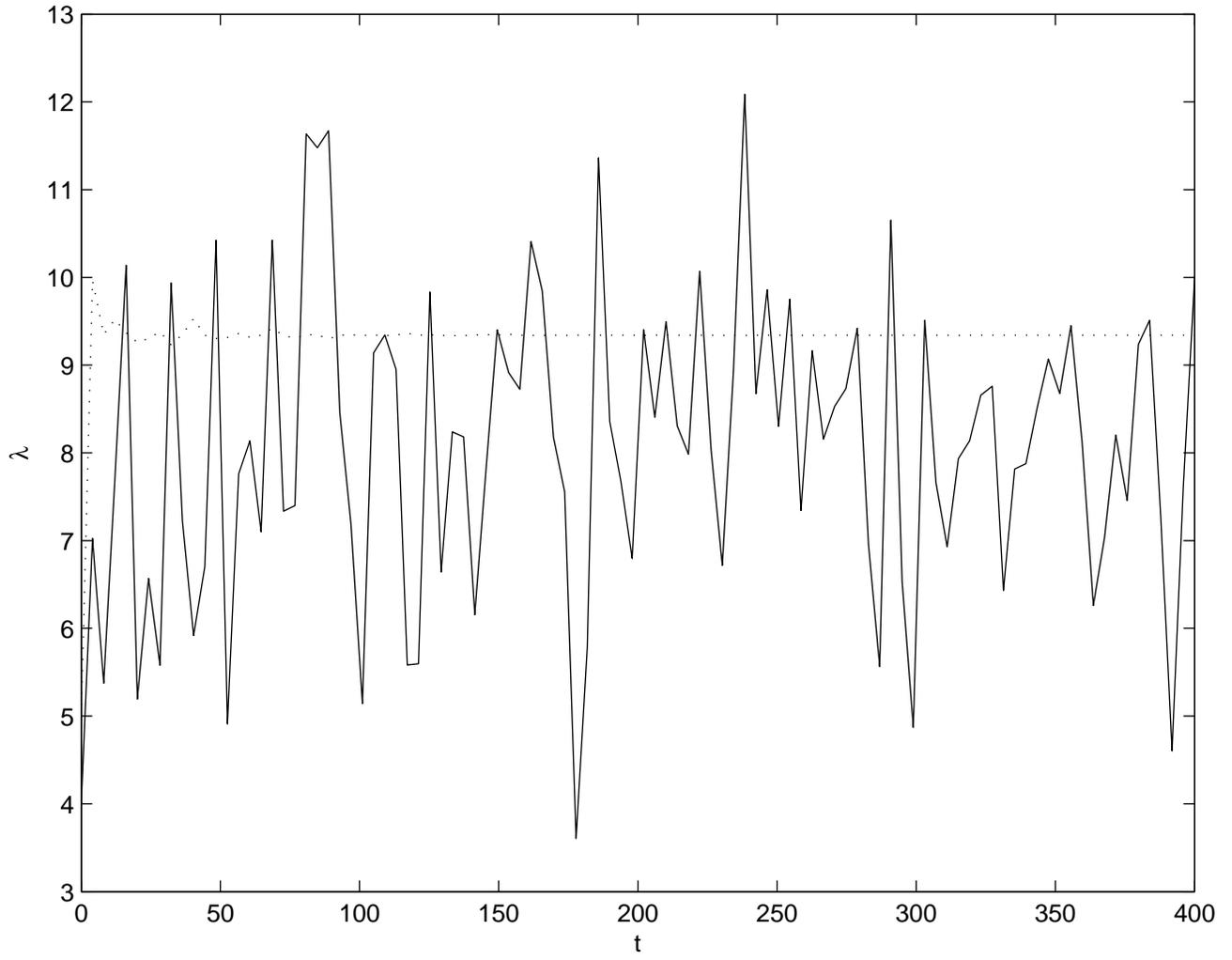
FIG.8. The wavelength-time plot (Eq. (15)) for the combined sinusoidal and chaotic time delay feedback: $\tau (t)=3 + 2\lambda(t)\sin 10t $. The time series of the wavelength for the variable (dotted line) and a constant (solid line) time delay are shown. Parameter values as in Fig.6. Dimensionless units. Combined sinusoidal and chaotic time delay feedback steers the chaotic wavelength dynamics to the stationary state.\\

\end{document}